\documentclass[aps,prl,notitlepage,10pt,twocolumn,superscriptaddress]{revtex4-1}
\usepackage{dcolumn}
\usepackage{bm}
\usepackage{amsmath,amsthm,amssymb}
\usepackage{epstopdf}
\usepackage{graphicx}
\usepackage{color}
\usepackage{verbatim}

\begin{document}
\title{Observation of Efimov Resonances in a Mixture with Extreme Mass Imbalance}

\author{R. Pires}\thanks{Both authors contributed equally to this work.}
\author{J. Ulmanis}\thanks{Both authors contributed equally to this work.}
\author{S. H\"afner}
\author{M. Repp}
\author{A. Arias}
\author{E. D. Kuhnle}
\affiliation{Physikalisches Institut, Universit\"at Heidelberg, Im Neuenheimer Feld 226, 69120 Heidelberg, Germany} 
\author{M. Weidem\"uller}\email{weidemueller@uni-heidelberg.de}
\affiliation{Physikalisches Institut, Universit\"at Heidelberg, Im Neuenheimer Feld 226, 69120 Heidelberg, Germany}
\affiliation{Hefei National Laboratory for Physical Sciences at the Microscale, University of Science and Technology of China, Hefei, Anhui 230026, PR China} 

\date{\today}

\begin{abstract}

We observe two consecutive heteronuclear Efimov resonances in an ultracold Li-Cs mixture by measuring three-body loss coefficients as a function of magnetic field near a Feshbach resonance. The first resonance is detected at a scattering length of $a_-^{(0)}=-320(10)~a_0$ corresponding to $\sim 7 (\sim 3)$ times the Li-Cs (Cs-Cs) van der Waals range. The second resonance appears at $5.8(1.0)~a_-^{(0)}$ close to the unitarity-limited regime at the sample temperature of $450$~nK. Indication of a third resonance is found in the atom loss spectra. The scaling of the resonance positions is close to the predicted universal scaling value of 4.9 for zero temperature. Deviations from universality might be caused by finite-range and temperature effects, as well as magnetic field dependent Cs-Cs interactions.
\end{abstract}
\maketitle

The control of interactions in ultracold atomic systems via magnetically tunable Feshbach resonances opens up new pathways for the investigation of few- and many-body physics~\cite{Chin2010}. One intriguing example is the access to the universal regime, which is characterized by a magnitude of the scattering length $a$ exceeding all other length scales of the system. In the limit of at least two resonant pairwise interactions, an infinite series of three-body bound-states, the so called Efimov states, exists~\cite{Efimov1971,Braaten2006,Zinner2013}. Counterintuitively, these trimers persist even for $a<0$, where the two body potential does not support a bound-state. The ratio between two subsequent trimer energies follows a discrete scale invariance with a universal scaling factor of $\mathrm{exp}(-2\pi/s_0)$. Here, $s_0$ only depends on the quantum statistics of the constituent atoms, their mass ratio, and the number of resonant interactions~\cite{DIncao2006,Braaten2006}. This scale invariance is also reflected in those values of $a$ where the energy of the bound-states coincides with the threshold of three free atoms for $a<0$, resulting in enhanced three-body loss. When the position of the first resonance is given by $a_-^{(0)}$, the $N-$th excited state is found at the scattering length $a_-^{(N)}=a_-^{(0)}\mathrm{exp}(\pi N/s_0)$. It has been shown that for homonuclear systems $a_-^{(0)}$ only depends on the characteristic range $r_0$ of the interatomic van der Waals potential~\cite{Berninger2011a,Roy2013,Wang2012a,Sorensen2012,Schmidt2012,Naidon2014}. The universal scaling factor acquires a value of 22.7 for equal mass constituents and features a drastic reduction in heteronuclear mass-imbalanced systems of two heavy and one light particle~\cite{Braaten2006,DIncao2006}, resulting e.g. in a factor of 4.9 for a $^6$Li-$^{133}$Cs mixture.

In ultracold atom experiments, Efimov resonances become evident in the three-body loss coefficient $L_3$ in the rate equation for atom loss $\dot{n}=-L_3 n^3$. Here, $n$ denotes the number density of atoms, and $L_3 \propto C(a) a^4$. The Efimov physics are contained in the dimensionless, log-periodic function $C(a)$. Thus far, Efimov resonances have been studied in several equal mass systems~ \cite{Kraemer2006,Gross2009,Ottenstein2008,Pollack2009,Wild2012,Huckans2009,Zaccanti2009,Berninger2011a,Roy2013}, where the scaling between different resonances is predicted to follow $C(a)=C(22.7 a)$. This large scaling factor demands a level of temperature and magnetic field control which makes the observation of an excited Efimov states highly involved. There had been indication of such an excited state in a three-component Fermi gas of $^6$Li\,atoms~\cite{Williams2009} exhibiting the same scaling as equal mass bosons~\cite{Braaten2010}. A finite temperature model~\cite{Rem2013} suggests that the experimental conditions of current experiments~\cite{Rem2013,Dyke2013} are close to temperature regimes where the observation of a second excited resonance in bosonic $^7$Li becomes feasible.

In heteronuclear systems, only K-Rb mixtures have been investigated so far~\cite{Barontini2009,Bloom2013}, where a scaling factor of $\sim 131$ obstructs the observation of an excited Efimov state. 
In $^6$Li-$^{133}$Cs the predicted scaling factor of 4.9~\cite{Braaten2006,DIncao2006} and the ability to tune the scattering length over a large range due to broad Feshbach resonances ~\cite{Repp2013,Tung2013} benefit the observation of a series of Efimov resonances.

In this Letter we present the observation of two consecutive Efimov resonances near the broad Feshbach resonance at 843 G in the energetically lowest  $^6$Li-$^{133}$Cs channel via measurements of the three-body loss coefficient. The assigned scattering lengths $a_-^{(0)}=-320(10)~a_0$ and $a_-^{(1)}=-1871(388)~a_0$ yield a ratio of $a_-^{(1)}/a_-^{(0)}=5.8(1.0)$, which is consistent with the ratio for two consecutive bound-states in a zero temperature model~\cite{Braaten2006,DIncao2006}. However, our analysis shows that for our experimental conditions the first excited Efimov state is already close to the unitarity limit, which leads to broadening of the resonant feature in the three-body loss spectrum and might additionally cause shifts of its position. Close to the anticipated position of the doubly excited Efimov state we observe another loss feature, but it is at the limit of our sensitivity of three-body loss coefficient measurements due to the unitarity-limited regime.

Our experimental setup is similar to the one presented in Ref.~\cite{Repp2013}. In brief, we load Cs into a crossed optical dipole trap (reservoir trap) at a wavelength of 1064~nm with 1/$e^2$ beam waists of 300~$\mu$m and a crossing angle of 90$^\circ$. We perform degenerate Raman sideband cooling~\cite{Treutlein2001} in order to prepare the majority of the atoms in the energetically lowest $\left| F=3,m_F=3\right\rangle $ spin state. Here, $F$ denotes the total atomic spin, and $m_F$ its projection. Lithium atoms in the ground state $\left| F=1/2\right\rangle $ manifold are loaded into another crossed dipole trap (dimple trap), located 1 mm away from the reservoir trap, which operates at a wavelength of 1070~nm, has 1/$e^2$ beams waists of 62~$\mu$m, and a crossing angle of $\sim 8^\circ$. 
After evaporative cooling of both species in the separated traps they are combined via a piezo driven mirror and the reservoir trap is subsequently turned off in a slow ramp. In the final evaporation step we sympathetically cool Cs close to the Li$\left| 1/2,-1/2\right\rangle \oplus\,$Cs$\left| 3,3\right\rangle$ Feshbach resonance at 943~G~\cite{Repp2013,Tung2013}, which expels all atoms in the Li$\left| 1/2,-1/2\right\rangle$ state from the trap. Finally, $1.6 \times 10^4$ (4$\times 10^4$) atoms remain in the  Cs$\left| 3,3\right\rangle$ (Li$\left| 1/2,1/2\right\rangle$) state. We measure trap frequencies 
$f_x, f_y, f_z$ 
of 114~Hz $\times$ 123~Hz $\times$ 11~Hz (275~Hz $\times$ 308~Hz $\times$ 33~Hz) and temperatures around 400~nK for both species. The magnetic fields are calibrated via microwave spectroscopy of Li, with a resolution of 30mG, limited by a residual magnetic field gradient along the long axis of the cigar shaped trap. Day-to-day drifts on the order of 20~mG add to the systematic uncertainty of the magnetic fields.

\begin{figure}[t!]
\includegraphics[width=0.99\columnwidth]{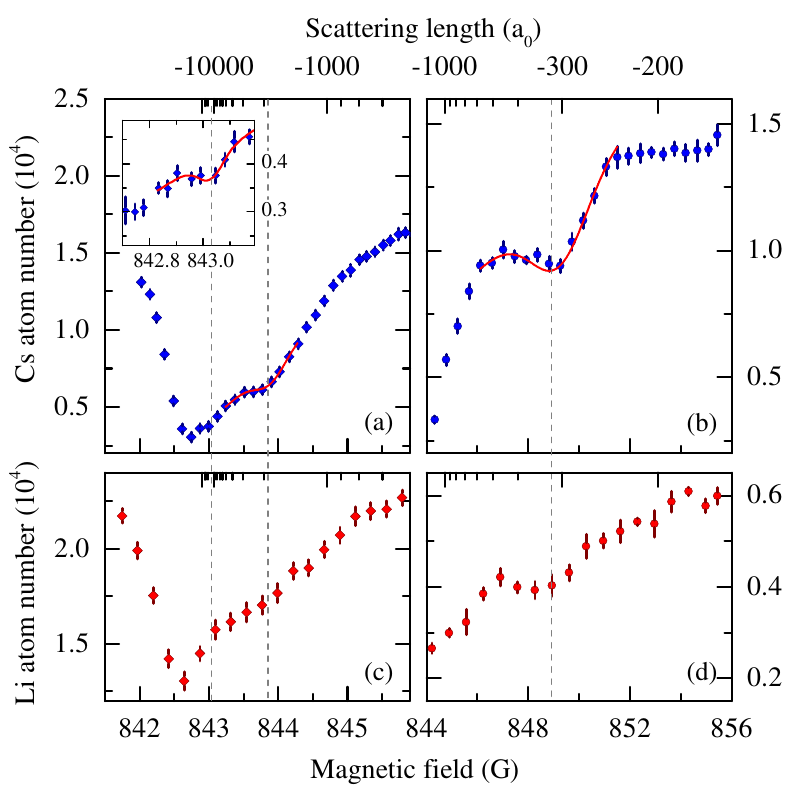}

\caption{Magnetic field dependent atom loss measurements of the Li-Cs mixture at a temperature of 400~nK. The Cs atom number after a hold time of 400 ms (a) and 1 s (b) is illustrated. The inset in (a) is a zoom into the region where a third loss feature can be seen. The Li atoms show resonances at consistent positions after a hold time of 1.2 s (c) and (d), when the initial Li atom number is reduced by a factor of two as compared to the measurements in (a) and (b). Each point is the mean of at least six independent measurements, and the error bars represent the standard error. The dashed vertical lines indicate the position of the Efimov resonances. The resonance positions are determined via a fit of Cs atoms with Gaussian profiles with linear background (red solid lines). The scattering length scale has been assigned via radio frequency association of universal dimers (see supplemental material).}
\label{fig:Efimov1}
\end{figure}

To measure the Efimov resonances, we ramp the magnetic field from 943~G to selected values near the broad Feshbach resonance at 843 G~\cite{Repp2013} and measure the Cs and Li atom number after a variable hold time by standard absorption imaging~\cite{Ketterle1999}. The results of our measurements are summarized in Fig.~\ref{fig:Efimov1}. The remaining number of Cs atoms for a hold time of 1 s is depicted in Fig.~\ref{fig:Efimov1}(b). To minimize the influence of systematic drifts we randomize the order in which the points are measured. As the increasing scattering length leads to fast losses closer to the Feshbach resonance, we perform a similar scan with a reduced hold time of 400 ms for the region closer to the Feshbach resonance, as shown in Fig.~\ref{fig:Efimov1}(a). Besides rising three-body losses associated with the increasing interspecies scattering length, we observe two clear features in the Cs loss data (dashed vertical lines). Fitting Gaussian profiles with linear background yields $B_0=849.12(6)_{\mathrm{stat}}(3)_{\mathrm{sys}}$~G and $B_1=843.89(1)_{\mathrm{stat}}(3)_{\mathrm{sys}}$~G for the positions of the resonances, where the first error denotes the statistic uncertainty of the fit and the second error results from the systematic uncertainty of the absolute magnetic field value. We identify these as LiCsCs Efimov resonances. We also detect a slight increase of the atom losses at a field of $B_2=843.03(5)_{\mathrm{stat}}(3)_{\mathrm{sys}}$ G, where the first error contains the width of fitted the Gaussian profile. The loss feature, which indicates a possible third resonance, is shown more clearly in the inset in Fig.~\ref{fig:Efimov1}(a).

Under the experimental conditions of Fig.~\ref{fig:Efimov1}(a) and (b), features in the Li loss signal are too small to be discerned from fluctuations of the Li atom number due to experimental instabilities. However, by lowering the Li atom number, which increases the fraction of lost atoms, we recover the Efimov resonances at consistent magnetic fields, as depicted in Fig.~\ref{fig:Efimov1}(c) and (d).


In order to verify that the observed Efimov resonances are indeed caused by a three-body process with the same ratio of Cs to Li atoms involved, the three-body loss coefficient is measured by observing the time dependent atom losses for various magnetic fields. The dimple trap depth is ramped up by about 10\% to reduce 
residual evaporation, but this also increases the temperature to $\sim~450$~nK. At the start of the measurement we prepare a mixture of $2 \times 10^4$ ($3 \times 10^4$) Cs (Li) atoms. After typical hold times on the order of a few seconds we lose nearly all Cs atoms, while the number of Li atoms only reduces by approximately 30\%. The temperature of the mixture remains unaffected within the uncertainties of our temperature determination ($\sim 15$\%).

The evolution of $n_\mathrm{Cs}$ is given by the rate equation
\begin{equation}
\dot{n}_\mathrm{Cs}=-L_1^{\mathrm{Cs}}n_\mathrm{Cs}-L_{3}^{\mathrm{LiCsCs}}n_{\mathrm{Li}}n_\mathrm{Cs}^2-L_{3}^{\mathrm{Cs}}n_\mathrm{Cs}^3.
\label{eq:lossrate}
\end{equation}
Here, $L_1^{\mathrm{Cs}}$, $L_{3}^{\mathrm{LiCsCs}}$ and $L_{3}^{\mathrm{Cs}}$ are the loss coefficients for Cs background collisions, Li+Cs+Cs three-body collisions, and Cs+Cs+Cs three-body collisions, respectively. The inter- and intraspecies two-body losses are ignored in Eq.~\eqref{eq:lossrate} because the atoms are in the energetically lowest states and thus only exhibit elastic two-body collisions. Under the conditions of the experiment the temperature dependence of the three-body loss coefficients~\cite{Rem2013} can be neglected, and $n_\mathrm{Li}$ can be assumed constant to a good approximation. Because the temperature is nearly constant, the change in density can be directly linked to the change in atom number.

Our analysis shows that the Cs atom loss curves are well described by Li+Cs+Cs and Cs+Cs+Cs losses, while a description assuming Li+Li+Cs and Cs+Cs+Cs losses alone does not reproduce the shape of the loss curves (see supplemental material). This verifies that Li+Li+Cs three-body losses are indeed strongly suppressed due to Fermi statistics and confirms that the observed Efimov resonances originate from the Li+Cs+Cs channel, which is the only channel for this mixture that is predicted to support universal three-body bound states~\cite{Braaten2006,DIncao2006}. This is also reflected in the loss ratio of Li to Cs atom numbers of $\sim$ 1:2 in the entire range of the magnetic fields probed.

In order to extract the loss rates from our measurements, we integrate Eq.~\eqref{eq:lossrate} over the spatial coordinates and fit the solution to the time dependent Cs atom number loss curves with $L_{3}^{\mathrm{LiCsCs}}$ as the only fitting parameter. $L_1^{\mathrm{Cs}}$ and $L_{3}^{\mathrm{Cs}}$ are independently obtained from single species measurements, and therefore do not enter as fitting parameters. We note that a close-by intraspecies Efimov resonance in Cs~\cite{Berninger2013} at 853~G does not influence determination of the Li-Cs Efimov resonances, since $L_{3}^{\mathrm{Cs}}$ is approximately constant in its direct vicinity and changes only in the region $a_{\mathrm{LiCs}}>-250~a_0$ (see supplemental material).

\begin{figure}[t!]
\includegraphics[width=0.99\columnwidth]{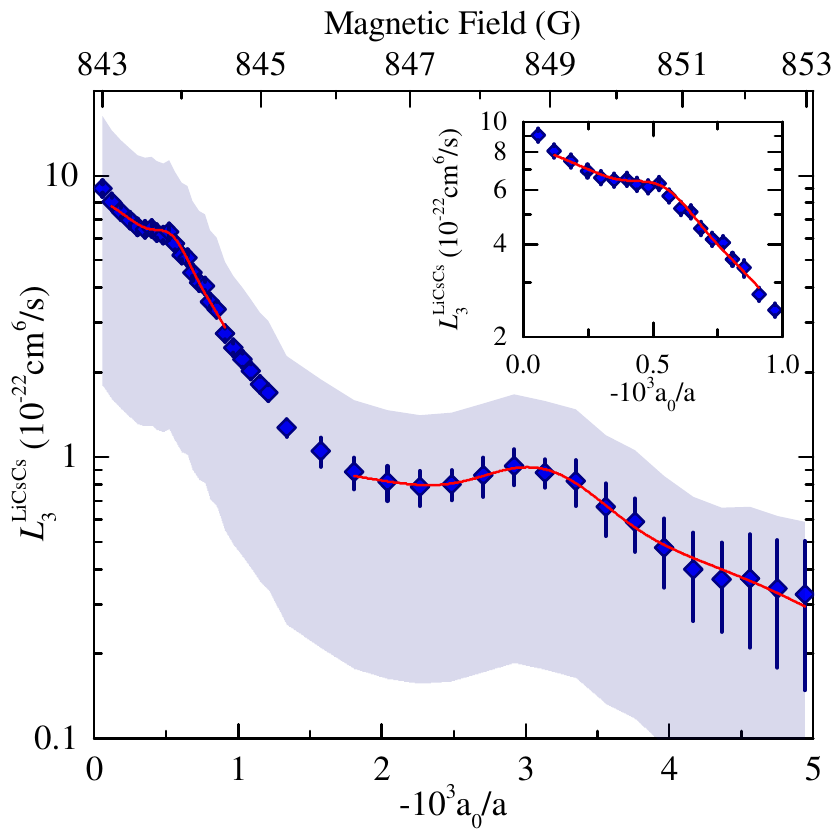}
\caption{Three-body loss coefficient $L_3^{\mathrm{LiCsCs}}$ plotted versus the inverse scattering length $1/a$. The blue diamonds show the mean of three $L_3^{\mathrm{LiCsCs}}$ measurements, where the error bars are given by the standard error.
The red solid lines show Gaussian profiles with linear background fitted to the data to determine the position of the two Efimov resonances. The grey area illustrates the systematic error of 80\% for the absolute value of $L_3^{\mathrm{LiCsCs}}$. The inset shows a zoom into the region of the first excited Efimov resonance.}\label{fig:Efimov2}
\end{figure}

The extracted three-body loss coefficient $L_3^{\mathrm{LiCsCs}}$ is depicted in Fig.~\ref{fig:Efimov2}. We estimate that the systematic error for the absolute value of $L_3^{\mathrm{LiCsCs}}$ is on the order of 80\%, due to uncertainties in measured atom numbers, temperatures, and trap frequencies. Additionally, the gravitational sag reduces the spatial overlap of the atom clouds. We estimate that this effect reduces the spatial integral over the densities in Eq.~\eqref{eq:lossrate} by $\sim 20$\%, and include this reduction into our model. Day-to-day drifts in the beam pointing of the dimple trap might cause fluctuations of the overlap.

A precise determination of the field dependent scattering length $a(B)$ is essential for a quantitative analysis. In particular the exact value of the scattering pole of the Feshbach resonance $B_{\mathrm{FR}}$, where $|a|$ diverges, plays a crucial role. We measure this value via radio frequency spectroscopy of universal dimers, which yields $B_{\mathrm{FR}}=842.9(2)$~G and $\Delta B =61.4(7)$\,G for the resonance position and width, respectively (see supplemental material). These values are in excellent agreement with an extensive study of Li-Cs Feshbach resonances via three different models~\cite{Pires}. Inserting these quantities into the relation
\begin{equation}
a(B)=a_{bg}\left( \frac{\Delta B}{B-B_{\mathrm{FR}}}+1 \right)
\label{eq:Scattering_length}
\end{equation}
(cf.~\cite{Moerdijk1995}) allows us to determine the abscissa in Fig.~\ref{fig:Efimov2}, where $a_{bg}=-28.5 a_0$~\cite{Repp2013}. 

We observe two distinct resonances in the $L_3^{\mathrm{LiCsCs}}$ measurements, which are consistent with the enhanced atom losses in Fig.~\ref{fig:Efimov1}. 
For large values of the scattering length, the loss rate approaches a value that is consistent with an order of magnitude estimate for the unitarity limit $L_3^{\mathrm{Lim}}\approx 10^{-21} \mathrm{cm}^6/\mathrm{s}$ for the temperatures in our experiments~\cite{Greene}.

Due to the lack of a finite temperature model for heteronuclear Efimov resonances similar to the one in Ref.~\cite{Rem2013}, the position of the resonances is determined via a fit of a Gaussian profile with linear background, which results in $B_0=848.90(6)_{\mathrm{stat}}(3)_{\mathrm{sys}}$~G and $B_1= 843.85(1)_{\mathrm{stat}}(3)_{\mathrm{sys}}$~G. Using Eq.~\eqref{eq:Scattering_length}, we assign the scattering lengths $a_-^{(0)}=-320(3)_{\mathrm{stat}}(2)_{\mathrm{sys}}(10)_{\mathrm{rf}}~a_0$ and  $a_-^{(1)}=-1871(19)_{\mathrm{stat}}(58)_{\mathrm{sys}}(388)_{\mathrm{rf}}~a_0$ for the Efimov resonance positions, where the third error accounts for the uncertainties in the Feshbach resonance position and width as extracted from the radio frequency spectroscopy (see supplemental material).

For the scaling between the first and second Efimov resonance positions, we obtain $a_-^{(1)}/a_-^{(0)}=5.8(0.1)_{\mathrm{stat}}(0.2)_{\mathrm{sys}}(1.0)_{\mathrm{rf}}$. This value is close to the predicted scaling of 4.9~\cite{Braaten2006,DIncao2006} for a zero temperature gas in the universal limit ($|a| \gg r_0$). 

The assumption of universal behavior is not strictly justified for the first resonance, since $a_-^{(0)}$ is only a factor of $\sim 7$ ($\sim 3$) larger than the Li-Cs (Cs-Cs) van der Waals length $r_0^{\mathrm{LiCs}}=45~a_0$ ($r_0^{\mathrm{Cs}}=101~a_0$), suggesting that finite range corrections might be important. The second resonance is already influenced by the unitarity limit at the temperatures achievable in the current experiment, which leads to a broadening of the resonance feature and might also cause additional shifts. 
In addition, the variation of the Cs intraspecies scattering length between -1200 $a_0$ at $B_0$ and -1500 $a_0$ at $B_1$~\cite{Berninger2013} still needs to be accounted for in the theoretical determination of the expected scaling ratio. Preliminary calculations~\cite{Wang} based on the model of Ref.~\cite{Wang2012b} suggest that the observed ratios are consistent with a slightly larger scaling factor. However, these results are subject to ongoing investigation and are beyond the scope of this report.

The position of the third resonance is indicated by a modulation of the atom losses at $B_2=843.03(5)_{\mathrm{stat}}(3)_{\mathrm{sys}}$\,G in Fig.~\ref{fig:Efimov1}, for which we assign the scattering length $a_-^{(2)}=-13.5(5.2)_{\mathrm{stat}}(3.1)_{\mathrm{sys}}\times 10^3~a_0$. The error due to systematic uncertainties in $B_{\mathrm{FR}}$ is on the order of $a_-^{(2)}$ and can even result in positive scattering lengths at $B_2$.
In a dedicated measurement of $L_3^{\mathrm{LiCsCs}}$ performed in the vicinity of $B_2$, we do not resolve an additional loss feature associated with the third resonance. However, this feature is in a regime where the scattering length is on the order of the thermal wavelength, and therefore not the dominating length scale, and a potential second excited Efimov resonance would be significantly reduced to the point where it can no longer be resolved. As a result, even though the losses at $B_2$ in Fig.~\ref{fig:Efimov1} are likely caused by Li+Cs+Cs collisions, we cannot unambiguously validate this hypothesis at this point.


After having found two consecutive Efimov resonances in the three-body loss rate coefficient of a mixture with large mass imbalance, the next experimental step will be the creation of a Li-Cs mixture at significantly lower temperatures. The unitarity-limited regime will be pushed towards larger values of the scattering length due to its scaling with $\propto 1/T^2$, and the resonances should become narrower and exhibit smaller shifts. In the current experiment, the gravitational sag due to the large mass of Cs leads to a separation of the atom clouds, which limits the lowest achievable temperatures of the mixture. Therefore, a dedicated engineering of trap potentials with species selective trapping forces~\cite{LeBlanc2007} will be  required. Based on a zero-range model~\cite{Helfrich2010} and calculations by C. Greene et al.~\cite{Wang} we estimate that temperatures on the order of 30~nK result in a unitarity-limited three-body loss coefficient of $\sim 2\times 10^{-19} \mathrm{cm}^6/s$, which is more than two orders of magnitude larger than the current limit. In order to avoid the influence of the unitarity limit on the third resonance, one would require temperatures on the order of $\sim 1$~nK.  Extending the model presented in Ref.~\cite{Rem2013} to the case of heteronuclear mixtures, however, might allow for a detailed analysis of the influence of unitarity at higher temperatures. An intriguing perspective is the study of finite size effects on the Efimov trimers~\cite{Portegies2011}, as the size of the doubly excited Efimov trimer of $\sim 0.2~\mu$m is of the same order as the oscillator length of the trapping potential used in the present experiments. A second major improvement over the current investigations will be a more precise determination of the magnetic field scattering pole $B_{\mathrm{FR}}$ in order to reduce the uncertainties in the assignment of scattering lengths to the observed resonances. Here, we expect an improvement by an order of magnitude will be possible by performing additional spectroscopy on the binding energy of the universal dimer with improved magnetic field stability. A more accurate determination of the resonance positions would shed new light on the application of universal few-body theories to mixed systems with large mass imbalance, addressing, e.g., to which extent the position of the first Efimov resonances also features universal scaling as recently found in homonuclear systems~\cite{Berninger2011a,Roy2013}.

Note: A recent preprint~\cite{Huang2014} reports the observation of two consecutive Efimov resonances in the system of three Cs atoms. Atomic loss features in Li-Cs under comparable experimental conditions are presented in the preprint~\cite{Tung2014}. 
The ratios between the first and second resonance are consistent within the errors. 
In contrast, we refrain from discussing the third resonance in terms of universal scaling due to the unitarity regime.  If we perform an analysis of the features in the loss spectra analogous to Ref.~\cite{Tung2014}, where a parabola is fitted to the global loss minimum to determine $B_{\mathrm{FR}}=842.73(1)_{\mathrm{stat}}(3)_{\mathrm{sys}}$,  we obtain $a_-^{(1)}/a_-^{(0)}=5.07(6)_{\mathrm{stat}}(13)_{\mathrm{sys}}(2)_{\mathrm{FR}}$ and $a_-^{(2)}/a_-^{(1)}=3.79(23)_{\mathrm{stat}}(39)_{\mathrm{sys}}(6)_{\mathrm{FR}}$ for the scaling ratios (errors labeled by $\mathrm{\mathrm{FR}}$ are representing the fit uncertainty of $B_{\mathrm{FR}}$).

\acknowledgements{This work is supported in part by the Heidelberg Center for Quantum Dynamics. R.P. and S.H. acknowledge support by the IMPRS-QD and J.U. by the DAAD. E.K. acknowledges support by the Baden-W\"urttemberg Stiftung. We thank R. Heck, A. Sch\"onhals and C. Renner for contributions to the experimental apparatus and Z. Pan for the initial calculations of the atom loss curves. We are grateful to R. Grimm, S. Jochim, P. Julienne, C. Salomon, and S. Whitlock for fruitful discussions, and E. Tiemann, J. Wang, J. D'Incao and C. Greene for insightful discussions and sharing their calculations with us.}

\bibliography{Efimov}

\end{document}


\title{Supplemental Material for \textquotedblleft Observation of Efimov Resonances in a Mixture with Extreme Mass Imbalance\textquotedblright}

\author{R. Pires}\thanks{Both authors contributed equally to this work.}
\author{J. Ulmanis}\thanks{Both authors contributed equally to this work.}
\author{S. H\"afner}
\author{M. Repp}
\author{A. Arias}
\author{E. D. Kuhnle}
\affiliation{Physikalisches Institut, Universit\"at Heidelberg, Im Neuenheimer Feld 226, 69120 Heidelberg, Germany} 
\author{M. Weidem\"uller}\email{weidemueller@uni-heidelberg.de}
\affiliation{Physikalisches Institut, Universit\"at Heidelberg, Im Neuenheimer Feld 226, 69120 Heidelberg, Germany}
\affiliation{Hefei National Laboratory for Physical Sciences at the Microscale, University of Science and Technology of China, Hefei, Anhui 230026, PR China} 

\pacs{}
\date{\today}

\maketitle


\subsection{Three-body losses of Cs atoms}
\begin{figure}[b!]
\includegraphics[width=0.99\columnwidth]{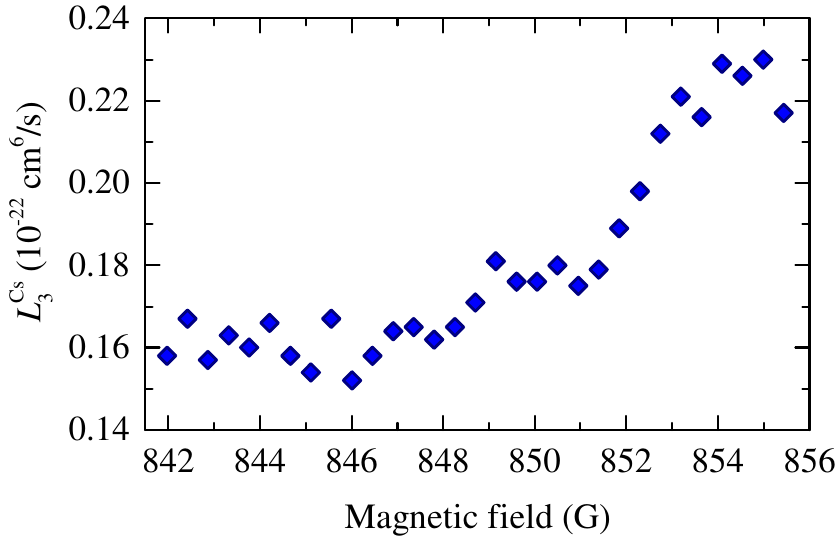}
\caption{Intraspecies three-body loss coefficient $L_3^{\mathrm{Cs}}$ of Cs atoms. Each point represents a single measurement of the $L_3^{\mathrm{Cs}}$ coefficient at a temperature of 410 nK.}
\label{fig:Supp:L3}
\end{figure}

In order to reduce the number of free parameters for the extraction of  $L_3^{\mathrm{LiCsCs}}$ from our loss curves via the rate equation
\begin{equation}
\dot{n}_\mathrm{Cs}=-L_1^{\mathrm{Cs}}n_\mathrm{Cs}-L_{3}^{\mathrm{LiCsCs}}n_{\mathrm{Li}}n_\mathrm{Cs}^2-L_{3}^{\mathrm{Cs}}n_\mathrm{Cs}^3,
\label{eq:lossrate}
\end{equation}
we independently measure the Cs intraspecies three-body loss coefficient $L_3^{\mathrm{Cs}}$ in a single species experiment. We observe the time dependent Cs atom losses at 410 nK temperature and extract the $L_3^{\mathrm{Cs}}$ coefficient from the spatially integrated Eq.~\eqref{eq:lossrate}, with $L_3^{\mathrm{LiCsCs}}=0$. The resulting rate dependence on the magnetic field is depicted in Fig.~\ref{fig:Supp:L3}. The $L_3^{\mathrm{Cs}}$ is approximately constant for the magnetic field values where the Li-Cs interspecies Efimov resonances are observed, and exhibits a maximum around 854 G due to the Efimov resonance in Cs~\cite{Berninger2011a}. For our analysis of $L_3^{\mathrm{LiCsCs}}$ via the full Eq.~\eqref{eq:lossrate}, we employ the average $L_3^{\mathrm{Cs}}$ value in the range from 842 G to 852 G as a fixed parameter.

Within the experimental limitations, these measurements were performed under the same conditions as the ones for the determination of $L_3^{\mathrm{LiCsCs}}$. Minor day-to-day drifts in the beam pointing of the dimple trap that might cause fluctuations of the trapping frequencies and trap depth add to the systematic error of the atom density. Based on the model from \cite{Rem2013} and \cite{Huang}, we estimate that the offset of the absolute $L_3^{\mathrm{Cs}}$ value due to the slightly lower temperature as compared to the Li-Cs interspecies experiments is well within our total systematic error, which is on the order of 80\%.

\subsection{Comparison of the loss channels}\
\begin{figure}[t!]
\includegraphics[width=0.99\columnwidth]{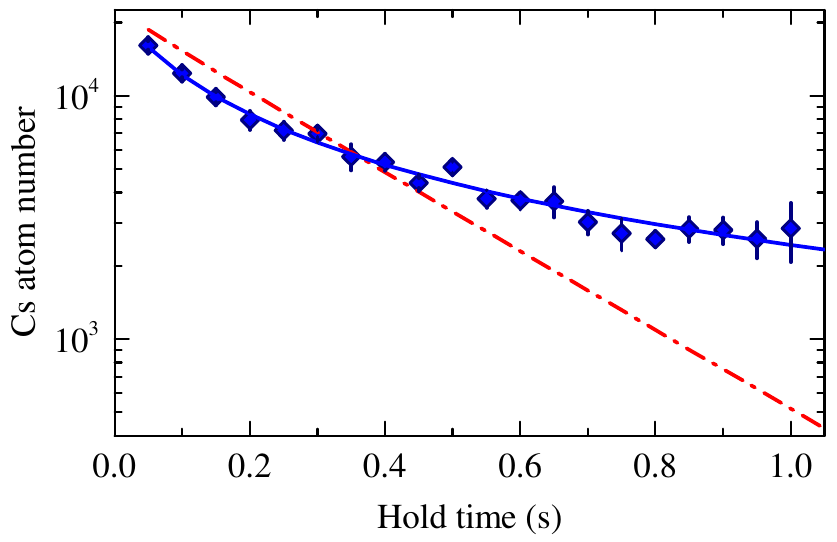}
\caption{Time-dependent atom losses in a Li-Cs \ mixture at a magnetic field of $843.9$~G. The data points show the Cs atom number, where the error is given by the standard error from at least five independent measurements. The blue solid line shows the results of a numerical fit with the spatially integrated Eq.~\eqref{eq:lossrate}, where Li+Cs+Cs and Cs+Cs+Cs three-body recombination are the dominating loss mechanisms. The red dash-dotted line demonstrates a numerical fit with  the spatially integrated Eq.~\eqref{eq:lossrate2}, where the Li+Cs+Cs loss channel is replaced by the Li+Li+Cs one.}
\label{fig:Supp:DecayCurves}
\end{figure}
We identify the dominating three-body recombination process by fitting the time-dependent Cs atom loss curves with the numerical solutions of the spatially integrated Eq.~\eqref{eq:lossrate}, where Li+Cs+Cs collisions are the dominating cause of losses. The fitted loss curve describes our experimental observations excellently, as shown in Fig.~\ref{fig:Supp:DecayCurves}. For comparison, we perform an analogous fit to the rate equation
\begin{equation}
\dot{n}_\mathrm{Cs}=-L_1^{\mathrm{Cs}}n_\mathrm{Cs}-L_{3}^{\mathrm{LiLiCs}}n_{\mathrm{Li}}^2n_\mathrm{Cs}-L_{3}^{\mathrm{Cs}}n_\mathrm{Cs}^3,
\label{eq:lossrate2}
\end{equation}
where Li+Li+Cs collisions are assumed as the dominating loss process. The result, as illustrated in Fig.~\ref{fig:Supp:DecayCurves}, clearly demonstrates that our data is not well described by this loss mechanism. This behavior is expected, since Pauli blocking prevents the two Li atoms in the same spin state to reach the small interparticle spacings that are necessary for three-body recombination to occur. Based on this, we neglect Li+Li+Cs losses in the formula for the time-dependent atom losses, given in Eq.~\eqref{eq:lossrate}.

\subsection{Position of the Feshbach resonance}\

To determine the pole of the Feshbach resonance we perform radio-frequency association measurements of the universal dimer binding energy. We start by creating a Li-Cs mixture in a procedure similar to the one presented in Ref.~\cite{Repp2013}. After combining both species in an optical dipole trap, we prepare all Li atoms in the $\left|1/2,-1/2\right\rangle$ spin state by illuminating the cloud with a resonant light pulse that expels all the atoms in the $\left|1/2,1/2\right\rangle$ spin state from the trap. The Cs spin state is prepared during the degenerate Raman sideband cooling, which polarizes most of the Cs atoms in the $\left| 3,3\right\rangle$ spin state. At this point we typically are left with $1.3 \times 10^4$ ($3 \times 10^4$) Cs (Li) atoms at 1.0 (0.5) $\mu$K temperature. 

\begin{figure}[t!]
\includegraphics[width=0.99\columnwidth]{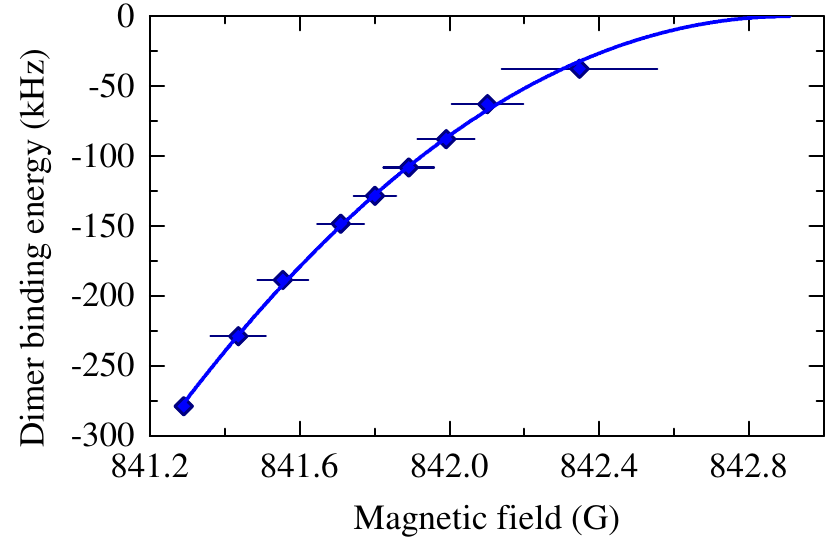}
\caption{Binding energy of the Li-Cs Feshbach dimer. The error bars represent the FWHM of a Gaussian profile fit to each of the loss features. The blue solid line shows the fit of the universal binding energy, which results in $B_{\mathrm{FR}}=842.90(3)$ G and $\Delta B=61.4(7)$~G. }
\label{fig:Supp:dimer}
\end{figure}

We associate the universal dimers by driving the system with a radio-frequency (rf) pulse with a frequency that is close to the resonance frequency between the two energetically lowest Li spin states. In order to determine the binding energy, we fix the rf frequency to a value slightly larger than the atomic resonance frequency and scan the magnetic field while observing the fraction of the Li atoms that are left in the initial state. The observed loss spectra are well described by Gaussian profiles, from which we extract the binding energies of the dimer. The results are summarized in  Fig.~\ref{fig:Supp:dimer}. The pole $B_{\mathrm{FR}}$ and width $\Delta B$ of the Feshbach resonance is determined by fitting the data with the universal dimer energy $E_b(B)=\hbar^2/\left(2\mu a(B)^2\right)$, where $\mu$ is the reduced mass, B represents the magnetic field, and $a(B)=a_{bg}\left(\Delta B/\left(B-B_{\mathrm{FR}}\right)+1 \right)$ with $a_{bg}=-28.5$~$a_0$~\cite{Repp2013}. The scattering length in the entire magnetic field range of these measurements is larger than 2500~$a_0$, and therefore corrections due to the van der Waals parameter $\bar{a}=42.8$~$a_0$ \cite{Chin2010} are neglected in the fit.  The dominating uncertainty arises from the stability of the magnetic fields, which at the time of the measurements was 170~mG. Finite range, temperature and density effects, which might affect the shape and position of the observed loss features, are dominated by the magnetic field instability and therefore neglected in this analysis.

The Feshbach resonance position obtained in this way differs from the previously reported one~\cite{Repp2013} for two main reasons. The previous determination relied on fitting a Gaussian profile to the loss feature, which did not take into account the asymmetric behavior of the scattering observables close to the resonance~\cite{tiecke2010}, as can be seen in Fig. 1 from the main Letter. Furthermore, magnetic field drifts on the order of 200~mG for the previous Feshbach resonance measurements~\cite{Repp2013} might add to the deviation.

\bibliography{Efimov}